\documentclass[aps,prb,a4paper,twocolumn,showpacs,showkeys,fleqn,floatfix]{revtex4-1}

\usepackage[dvips]{epsfig}
\usepackage[dvips]{color}
\usepackage{amsmath}

\newcommand{\figwidth}{0.9\columnwidth}
\newcommand{\sz}{\widehat{S}^z{}}
\newcommand{\sx}{\widehat{S}^x{}}
\newcommand{\sy}{\widehat{S}^y{}}

\newcommand{\svect}{\widehat{\mathbf{S}}}
\newcommand{\vect}[1]{\mathbf{#1}}
\newcommand{\ham}{\widehat{\cal{H}}}

\newcommand{\boldeta}{\boldsymbol\eta}

\begin{document}
\title{Modes of magnetic resonance of S=1 dimer chain compound NTENP.}

\author{V. N. Glazkov}
\email{glazkov@kapitza.ras.ru}

\affiliation{Kapitza Institute for Physical Problems, Kosygin str.
2, 119334 Moscow, Russia\\}

\author{A. I. Smirnov}

\affiliation{Kapitza Institute for Physical Problems, Kosygin str.
2, 119334 Moscow, Russia\\Moscow Institute for Physics and
Technology, 141700 Dolgoprudny, Russia}

\author{A.Zheludev}

\affiliation{ETH Z\"{u}rich, Laboratorium f\"{u}r
Festk\"{o}rperphysik,  8093 Z\"{u}rich, Switzerland}

\author{B.C. Sales}

\affiliation{Oak Ridge National Laboratory, Oak Ridge, Tennessee
37831-6393, USA}

\date{\today}

\begin{abstract}
The spin dynamics of a quasi one dimensional $S=1$ bond alternating
spin-gap antiferromagnet Ni(C$_9$H$_{24}$N$_4$)NO$_2$(ClO$_4$)
(abbreviated as NTENP) is studied by means of electron spin
resonance (ESR) technique. Five modes of ESR transitions are
observed and identified: transitions between singlet ground state
and excited triplet states, three modes of transitions between spin
sublevels of  collective triplet states and antiferromagnetic
resonance absorption in the field-induced antiferromagnetically
ordered phase. Singlet-triplet and intra-triplet modes demonstrate a
doublet structure which is due to two maxima in the density of
magnon states in the low-frequency range. A joint analysis of the
observed spectra and other experimental results allows to test the
applicability of the fermionic and bosonic models. We conclude that
the fermionic approach is more appropriate for the particular case
of NTENP.
\end{abstract}


\maketitle

\section{Introduction.}

Dielectric magnetic crystals with an antiferromagnetic exchange
interaction  usually demonstrate N\'{e}el ordering at some critical
temperature. Nevertheless, a number of antiferromagnetic systems do
not order at temperatures far below the Curie-Weiss temperature.
Examples are to be found among quasi one dimensional magnetic
structures, networks of coupled magnetic dimers, frustrated magnets
etc. Some of these systems have a gap of exchange origin in the
magnetic excitation spectrum. This gap stabilizes the
spin-disordered ground state against perturbations, such as
interactions between the spin chains, anisotropy, etc. Spin-gap
magnets remain in a disordered ("spin-liquid") state down to zero
temperature. The energy gap in the spectrum of triplet excitations
may be varied by applying magnetic field or pressure. This provides
an opportunity  to control the stability of the spin-liquid state.
When the gap vanishes at some critical field, the spins order
antiferromagnetically. This phenomenon has been observed and studied
recently for different spin-gap systems, such as the dimer magnet
TlCuCl$_3$ ~\cite{tlcucl3-oosawa,tlcucl3-glazkov}, the Haldane
magnet PbNi$_2$V$_2$O$_5$ ~\cite{pbnivo-tsujii,pbnivo-smirnov}, the
spin $S$=1 anisotropic chain system NiCl$_2$-4SC(NH$_2$)$_2$
(abbreviated as DTN) \cite{DTNfirst,DTN}. Field induced
antiferromagnetic ordering in spin-liquid systems was intensively
discussed, in particular,  as a kind of Bose-Einstein condensation
of magnons (see, e.g. Refs.
\onlinecite{tlcucl3-nikuni},\onlinecite{giamarchi-nature}).

Elementary excitations of a Heisenberg spin-gap magnet are triplets,
which may be classified by their spin projection $S_z$ and momentum
$\mathbf{k}$. At zero field the lowest excited state is  three-fold
degenerate ($S_z=\pm1;0$). The energy of the spin sublevels
$S_z=\pm1$ depend on the magnetic field: $E(S_z=\pm1)=\Delta\mp
g\mu_BH$. The energy gap vanishes at a critical field
$H_c=\Delta/g\mu_B$. Relativistic terms (e.g., crystal field,
dipole-dipole interaction or anisotropic exchange) violate the high
symmetry of the Heisenberg Hamiltonian.  This results in a mixing of
different spin states and in the lifting of the zero-field
degeneracy of triplet spin sublevels. In this case the field
dependence of spin sublevels, and hence the relation between the
zero field energy and the critical field, become more complicated.
There are three approaches  to describing the spectrum in a magnetic
field and calculating the critical field: fermionic models
\cite{tsvelik,affleck}, the bosonic (or macroscopic) approach
\cite{affleck,farmar}, and perturbative calculations
\cite{zaliznyak,golinelly}. The fermionic approach is justified for
the case of a purely one-dimensional spin-gap system. The bosonic
approach implies proximity of  three dimensional magnetic order. It
is valid if the energy gap is much smaller then the main exchange
integral. The perturbative approach is valid at low fields $g\mu_B
H\ll\Delta$ irrespectively to the system dimensionality. It
demonstrates that quantized collective excitations are equivalent to
sublevels of a single spin $S=1$ in an effective crystal field.
Predictions of the perturbative model coincide with those of
fermionic model (see Ref.\onlinecite{zaliznyak}), thus expanding
applicability of the perturbative model in the one-dimensional case
up to critical field. If the zero field spin sublevels are fixed,
both types of results are indistinguishable at low field.
Nevertheless, they differ in the vicinity of critical field: the
perturbative/fermionic approach predicts a closing of the gap
linearly with the magnetic field, while macroscopic approach
predicts a faster square root dependence. The scenario of the spin
gap behaviour in a magnetic field is not universal: the Haldane
compound NENP was found to fit better to the fermionic model
\cite{affleck}, while another Haldane magnet PbNi$_2$V$_2$O$_8$  and
the dimer-type spin-liquid magnet TlCuCl$_3$ fit to the macroscopic
model \cite{pbnivo-smirnov, farmar}. Thus, it is of importance to
study further spin-gap systems which demonstrate closing of a spin
gap in a magnetic field.

Another problem that we address in this paper is the stability of
excitations in spin liquid systems. The magnetic excitations
spectrum usually includes several branches which may form
two-particle continua. A decay of magnons is possible, if a
single-magnon branch enters such a continuum. In this case magnon
modes become overdamped, often resulting in a termination of the
single-particle spectrum.

The recently found compound  Ni(C$_9$H$_{24}$N$_4$)NO$_2$(ClO$_4$),
abbreviated as NTENP, is a good test probe for spin excitations at a
critical field closing the  gap, as well as for field induced
ordering and spectrum
termination.\cite{ntenp-first,narumi,hagiwara-prl94,zheludev2004,zheludev2006}
Besides, it belongs to a new family of spin-gap magnets with
dimerised $S=1$ spin chains. In crystals of NTENP, Ni$^{2+}$
magnetic ions  form dimerised spin S=1 chains running along $a$ axis
of the triclinic crystal. In-chain exchange integral takes  two
alternating values $J_1$ and $J_2$ with a ratio
$\alpha=J_1/J_2=0.45$. The larger exchange integral is
$J_2/k_B=54.2$~K. \cite{narumi} The spin gap in $S=1$ chains may be
of a Haldane type, if chains are uniform ($\alpha=1$), or of a dimer
nature in the opposite limit case $\alpha=0$. There is a zero gap
point between these limiting cases at $\alpha\simeq$0.6 (see, e.g.,
Ref.~\onlinecite{roma2003}), thus NTENP belongs to the dimer class.
The ground state of NTENP is a nonmagnetic singlet separated from
the triplet excitations by an energy gap of about 1 meV. The triplet
energy levels are split by the crystal field: the energies of the
sublevels, found in neutron scattering experiments are 1.06, 1.15
and 1.96 meV, \cite{hagiwara-prl94} respectively. In zero magnetic
field, the doubled energy  of the lowest sublevel is a little bit
lower than the energy of the upper sublevel.  The excitations
energies are field-dependent, and in a magnetic field the decay of
magnons becomes allowed by the energy conservation
law.\cite{hagiwara-prl94,zheludev2006} The critical field $H_c$,
closing the spin-gap is anisotropic: it varies from 85 kOe in a
magnetic field applied along the chains to 114 kOe in the
perpendicular orientation. \cite{narumi-prb} The field-induced
ordering takes place only at low temperatures: in a magnetic field
of 100 kOe, applied along the chains, the N\'{e}el temperature is
about 0.5 K. The phase diagrams for the induced ordering at
different orientations of the magnetic field are described in
Refs.~\onlinecite{hagiwara-prl96}, \onlinecite{TateiwaPhysicaB}.
Antiferromagnetic ordering above $H_c$ is confirmed by means of
neutron scattering. \cite{hagiwara-prl94}

In the present paper we report the detailed electron spin resonance
(ESR) study of this material. This technique allowed us to detect
excitations with energies of about 0.05meV (corresponding to the
resonance frequency of approximately 10GHz) with the resolution of
about 0.005meV (1 GHz). We observed magnetic resonance modes of
thermally activated triplet excitations and the over-gap
transitions, as well as antiferromagnetic resonance absorption in
the field-induced ordered phase. The splitting of triplet modes was
observed, which can be ascribed to the dispersion of the triplet
excitations with wave vectors along the direction of a weak exchange
(i.e. perpendicular to chains). Further, the spin resonance data
provided a possibility for comparison of fermionic (perturbative)
and bosonic (macroscopic) models. We found that our data favour the
fermionic description. We refined parameters of the effective
Hamiltonian of the triplet excitations and estimate the inter-chain
exchange integral. The effects, related two the two-magnons decay of
spin excitations, were observed.

\section{Experimental details and samples.}

 A set of custom-built  transmission type ESR spectrometers covering
the frequency range from 9 to 260 GHz was used for the experiments.
The spectrometers were inserted  in a $^4$He-bath cryostats equipped
with 60---120 kOe cryomagnets. Microwave radiation  was produced by
commercial Gunn diode generators and backward wave oscillators. The
microwave resonator with the sample inside was placed in a
hermetically sealed chamber filled with heat exchange gas. The
temperature of the sample may be stabilized by means of a heater in
the interval between 1.3 K and 30 K. Experiments in the temperature
range 0.4---1 K were performed in a custom-built spectrometer
equipped with a $^3$He-3 pumping refrigerator. Microwave modes of
several resonators were used to cover the entire frequency range.

In an experiment the signal transmitted through the resonator is
recorded as a function of magnetic field. A diminishing of
transmission means microwave absorption. Resonance absorption
appears if the microwave frequency equals the frequency of a
transition between states $|\imath\rangle$, $|\jmath\rangle$
($\hbar\omega=|E_{\imath\jmath}|=|E_\imath-E_\jmath|$).  The
following selection rules should be satisfied: (i) the dipolar
transition matrix element
$\langle\imath|(\vect{h}\svect)|\jmath\rangle$ is non-zero (here
$\vect{h}$ is the amplitude of the oscillating field) and (ii)
states $|\imath\rangle$ and $|\jmath\rangle$ have the same
wavevector $\vect{k}$.  The total ESR absorption is a sum over all
pairs of spin states and an integral over the $\vect{k}$-space.

The configuration of the microwave field in the resonator is
different for different microwave modes, hence the polarization of
microwave magnetic field in the sample is also different. Due to the
finite sample size, even for a given mode, different polarizations
of microwave field are present within the sample. Therefore, for the
typical sample size, and for most of microwave resonator modes, the
polarization selection rules do not prohibit any modes of magnetic
resonance absorption.

We have used the samples of deuterated NTENP from the same batch as
in Ref.\onlinecite{zheludev2004}.

\section{Experimental results.}

\subsection{Above 1K: ESR of thermally activated excitations.}

\begin{figure}
  \epsfig{file=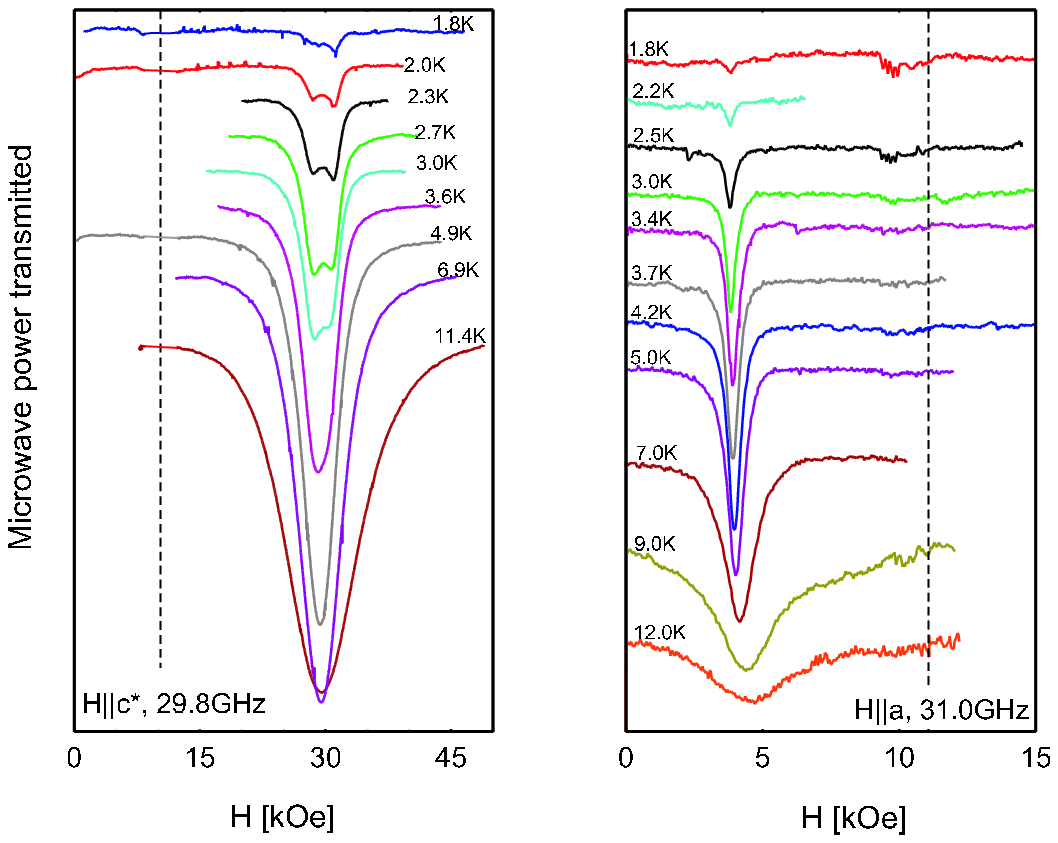, width=\figwidth, clip=}\\
  \caption{(color online) ESR absorption above 1K at
  $\vect{H}||c^*$ (left) and $\vect{H}||a$ (right).
  Vertical dashed lines  mark the position of the paramagnetic resonance with
  $g=2.00$.}\label{fig:scans(t)above1K}
\end{figure}

\begin{figure}
  \epsfig{file=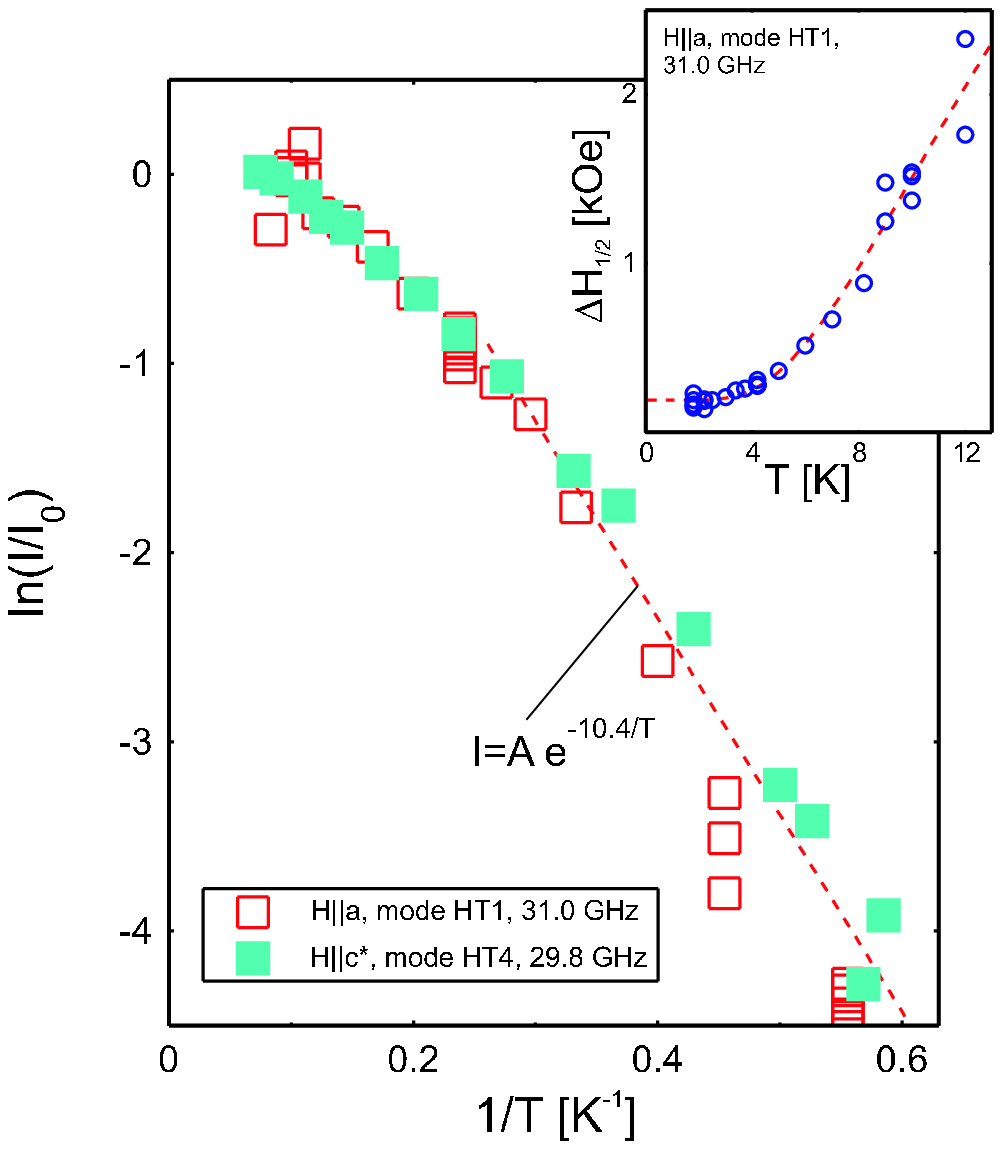, width=\figwidth, clip=}\\
  \caption{(color online) Temperature dependences of the integrated
intensity of ESR absorption. The dashed straight line on the main
panel corresponds to the exponential law $I=A\exp\{-\frac{E_0}{T}\}$
with $E_0=10.4$K. Inset: Temperature dependence of half-linewidth.
The dashed line at the inset is a guide to the eye.}\label{fig:i(t)}
\end{figure}

\begin{figure}
  \epsfig{file=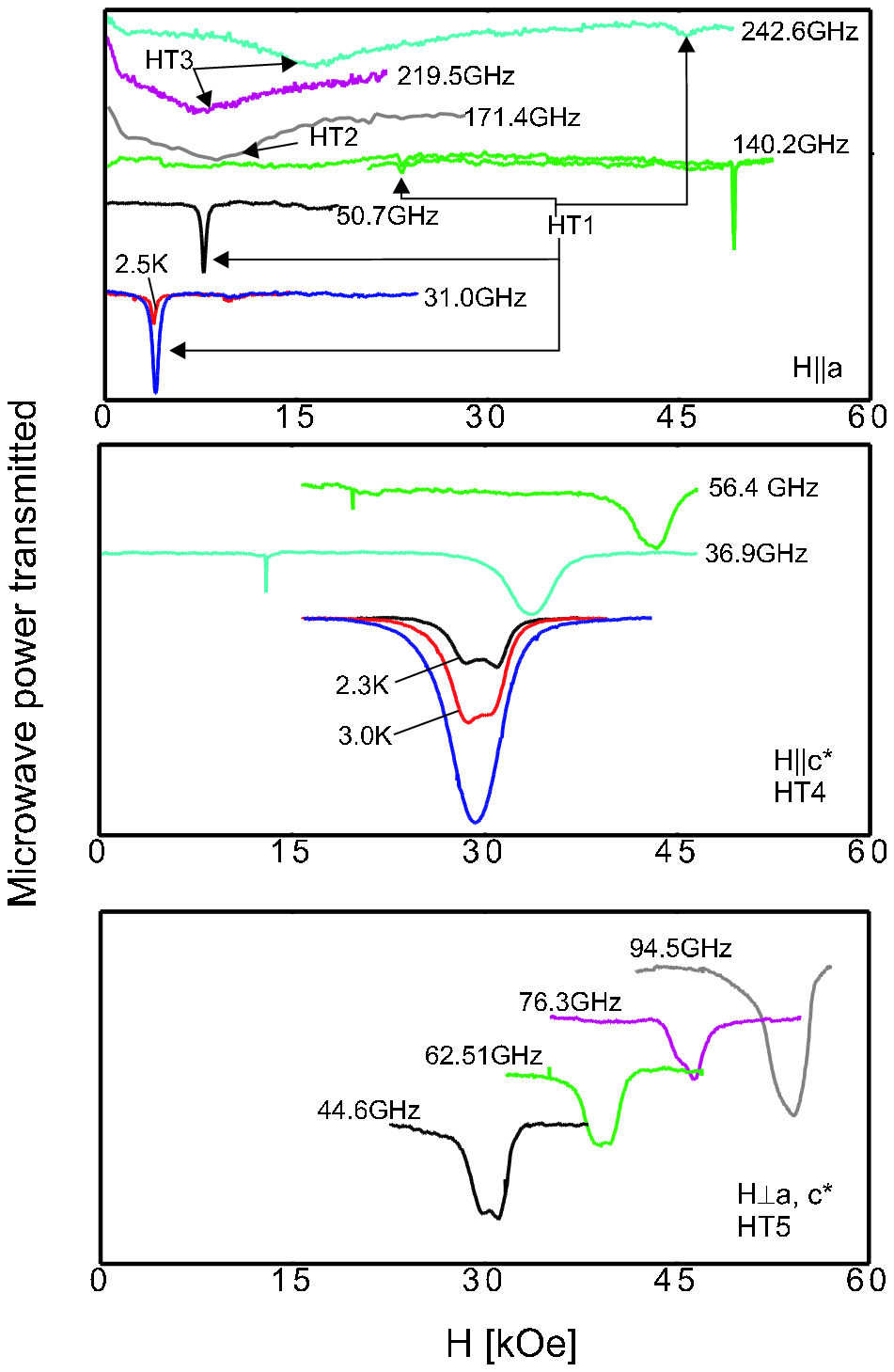, width=\figwidth, clip=}\\
  \caption{ (color online) ESR absorption lines for different resonance modes. All
  absorption curves, unless specified otherwise, are measured at
  4.2K. HT1...HT5  denote resonance modes as described in the text.
  The unlabelled narrow absorption line is a DPPH marker
  ($g=2.00$).}\label{fig:lineshapes}
\end{figure}

The intensity of the magnetic resonance absorption observed above 1K
increases with temperature up to about 20 K. Upon heating beyond 20
K the lines broaden and disappear. Examples of the temperature
evolution of spin resonance absorption for two orientations of the
magnetic field are presented in Fig. \ref{fig:scans(t)above1K}.  The
lines are strongly shifted from the paramagnetic resonance position
of free electron spins. The resonance field is anisotropic, which is
in agreement with the zero-field splitting of the triplet sublevels
observed in neutron scattering experiments. Below 5K the integrated
intensity follows a law of thermal activation $I\sim\exp\{-E_0/T\}$
with the activation energy $E_0=10.4\pm1.0$K (see Figure
\ref{fig:i(t)}). This activation energy corresponds well to the
known energy of lowest excited state of 1.06
meV.\cite{hagiwara-prl94} Exponential freezing out of intensity
proves that this signal is due to transitions between sublevels of
excited states. The temperature dependence of the linewidth is shown
on the inset of Figure \ref{fig:i(t)}). The linewidth  at $T$=4K is
practically the same as in the low-temperature limit. The resonance
field at 4K also coincides with its low-temperature value. Thus,
absorption at $T$=4K may be attributed to noninteracting excitations
distributed near the bottom of the spectrum.

ESR absorption lines were collected at  a set of frequencies at
$T$=4.2~K for three mutually orthogonal orientations: $\vect{H}||a$,
$\vect{H}||c^*$ and $\vect{H}\perp a, c^*$.  Here $a$ is the
crystallographic axis of the triclinic crystal and $c^*$ is
perpendicular to the crystallographic $(ab)$-plane. From here on we
will mark the resonance modes observed in the high temperature range
$T>$1 K as follows: HT1, HT2 and HT3 are the modes observed at
$\vect{H}||a$, numbered in the order of ascend of zero-field
frequency , HT4 is the mode observed at $\vect{H}||c^*$ and HT5 is
the mode observed at $\vect{H}\perp a,c^*$. Examples of the
resonance lines at different frequencies are given in Figure
\ref{fig:lineshapes}.

At $\vect{H}||a$ we observe single resonance lines for all modes.
For $\vect{H}||c^*$ or $\vect{H}\perp a,c^*$ at low temperature the
absorption line is split into two close components (see Figure
\ref{fig:lineshapes}). The intensities of the split components are
approximately equal. The high frequency modes HT2 and HT3, observed
in the $\vect{H}||a$ orientation, are much broader then other
resonance modes.

\subsection{Below 1K: Singlet-triplet transition and antiferromagnetic resonance.}

\begin{figure}
  \epsfig{file=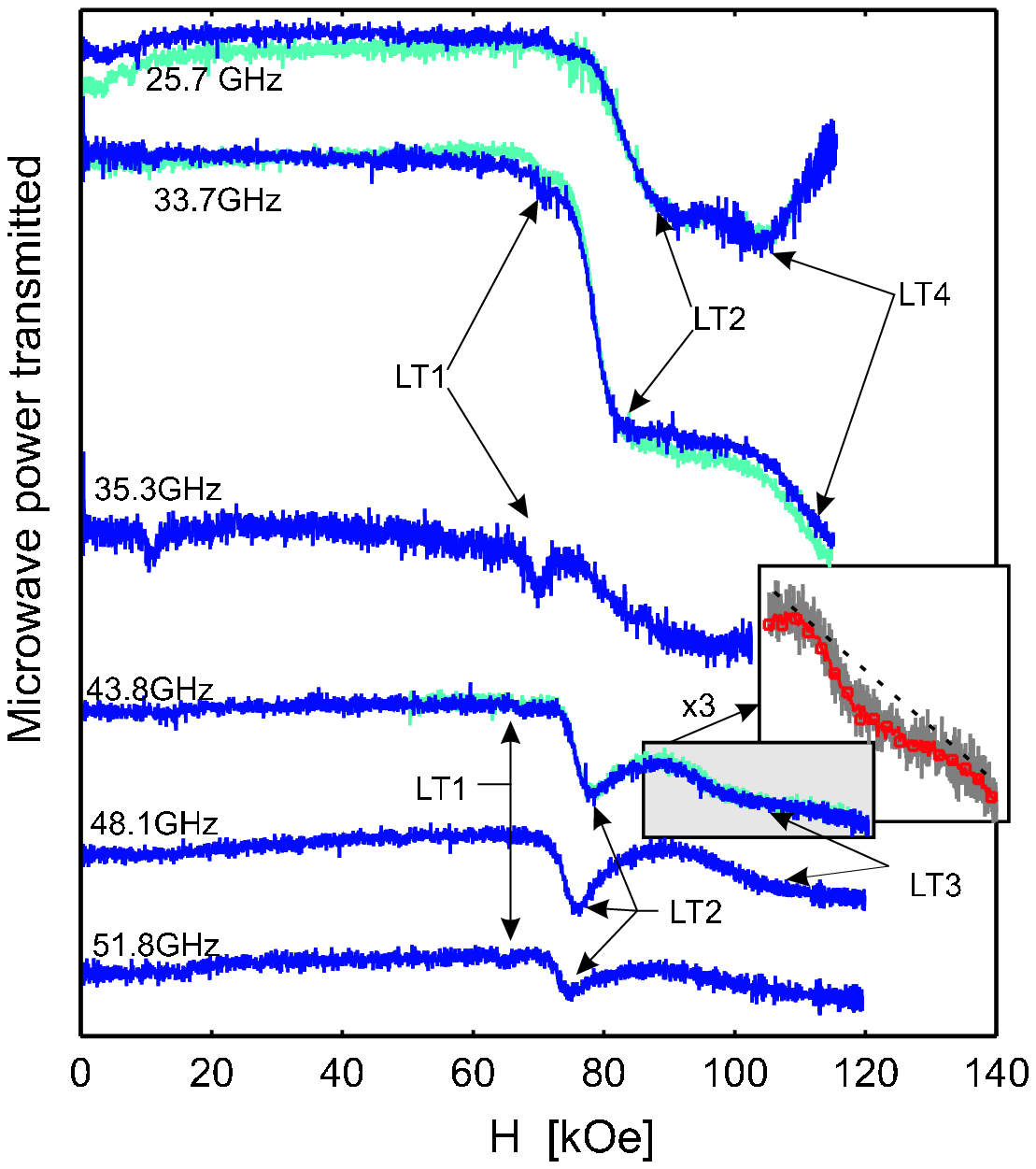, width=\figwidth, clip=}\\
  \caption{(color online) ESR absorption at 0.4K at different frequencies. LT1,
  LT2, LT3, LT4 are the absorption components (see
  text).$\vect{H}||a$}\label{fig:scans(f)below1K}
\end{figure}

\begin{figure}
  \epsfig{file=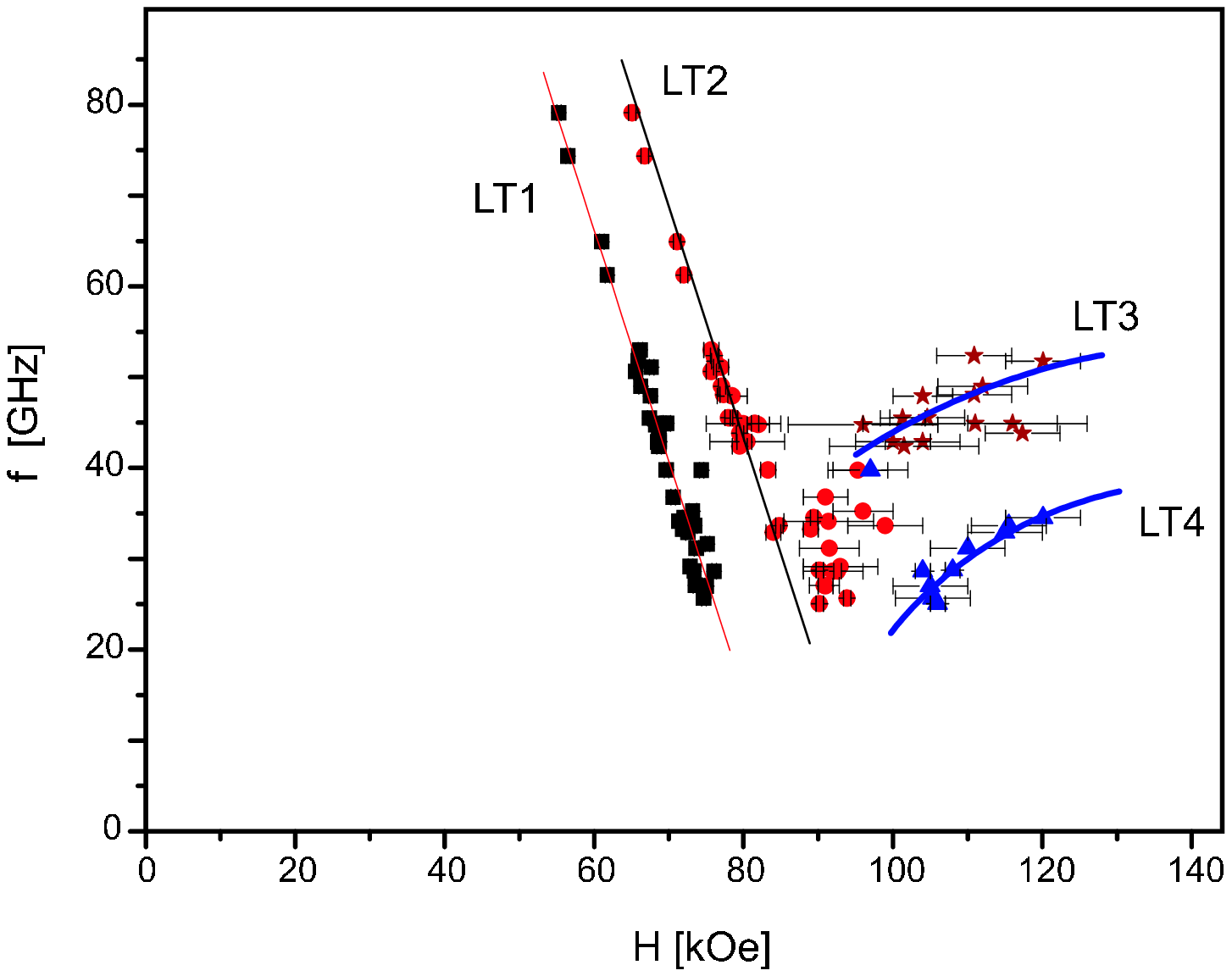, width=\figwidth, clip=}\\
  \caption{(color online) Frequency-field diagram for modes observed
  at  $T=0.4$K. Lines are guides to the eye. $\vect{H}||a$}
  \label{fig:LT3&LT4zoom}
\end{figure}

The experiments at $T<1$~K were performed only for $\vect{H}||a$
because the critical field $H_c$ is a minimum at this orientation,
enabling the largest value of $H/H_c$.

At temperatures below 1~K the triplet excitations are frozen out and
do not contribute to resonance absorption. Nevertheless, other types
of the ESR absorption arise at high fields (Figure
\ref{fig:scans(f)below1K}). First, two absorption components (LT1
and LT2) split by about 10 kOe are observed below 90 kOe. These
resonances shift to lower fields with the frequency increase. The
second low-temperature absorption range is found above 90 kOe. Here
two broad bands of microwave absorption denoted as LT3 and LT4 are
found. Both of the absorption maxima shift to the higher fields with
increasing frequency, as shown on Figure \ref{fig:LT3&LT4zoom}.

\begin{figure}
  \epsfig{file=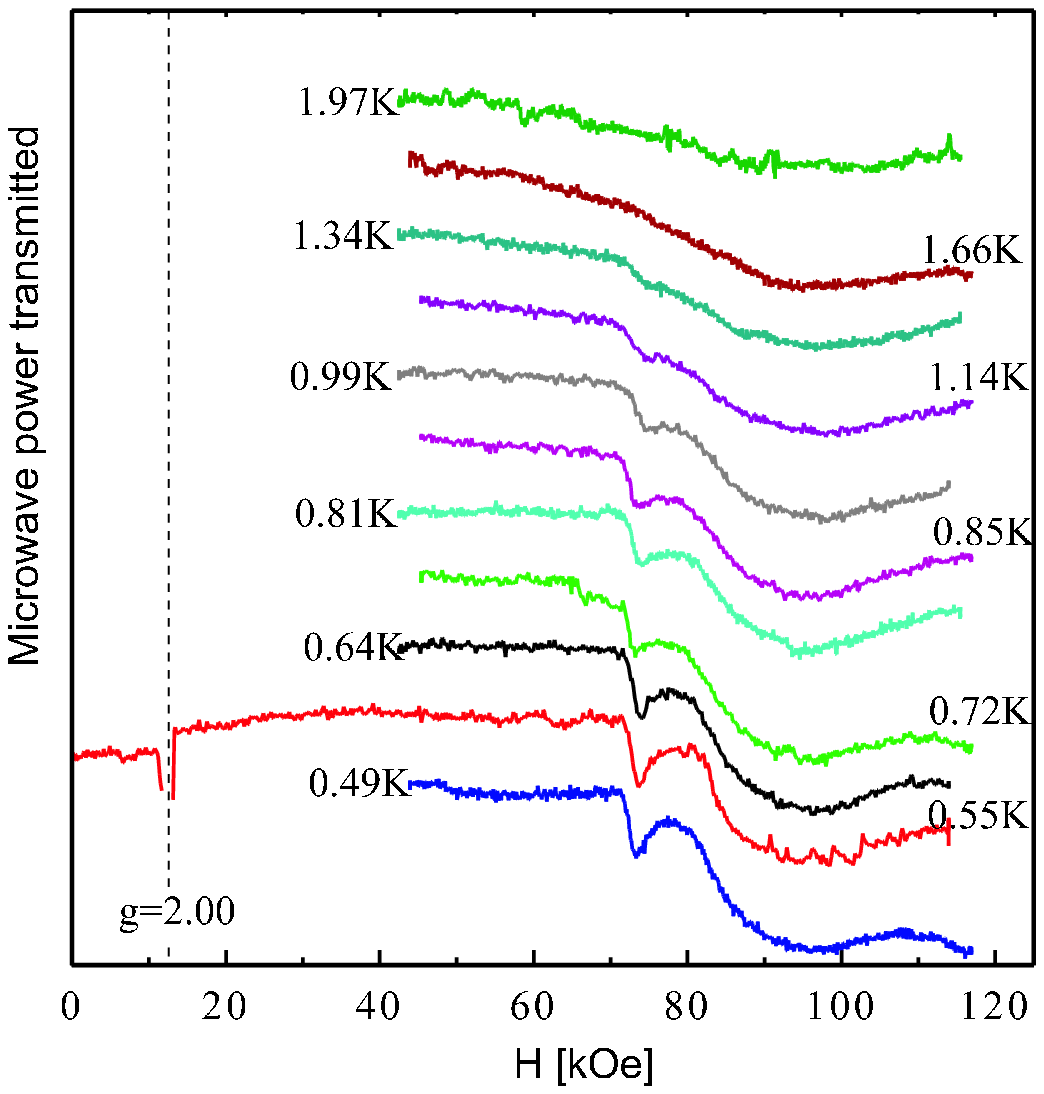, width=\figwidth, clip=}\\
  \caption{(color online) Temperature evolution of low-temperature modes at
  $f=$35.25 GHz, $\vect{H}\parallel a$. The vertical dashed line marks
  the position of a paramagnetic resonance with $g$=2.00.}
  \label{fig:35ghzvartatlowT}
\end{figure}

Mode LT1 is usually much weaker then LT2. The relative intensity of
these modes varies irregularly with the frequency. This behaviour
is, probably, due to the dependence of their excitation on the
dominating polarization of microwave field. In contrast to HT-modes,
intensities of modes LT1 and LT2 decrease with increasing
temperature (see Figure \ref{fig:35ghzvartatlowT}), i.e. they are
not thermally activated. These frequency-field and temperature
dependences allow us to conclude that modes LT1 and LT2 correspond
to singlet-triplet transitions. Two other low-temperature modes (LT3
and LT4) correspond to resonance modes of the field-induced ordered
phase (above H$_c$) and may be considered as antiferromagnetic
resonance modes.

\begin{figure}
  \epsfig{file=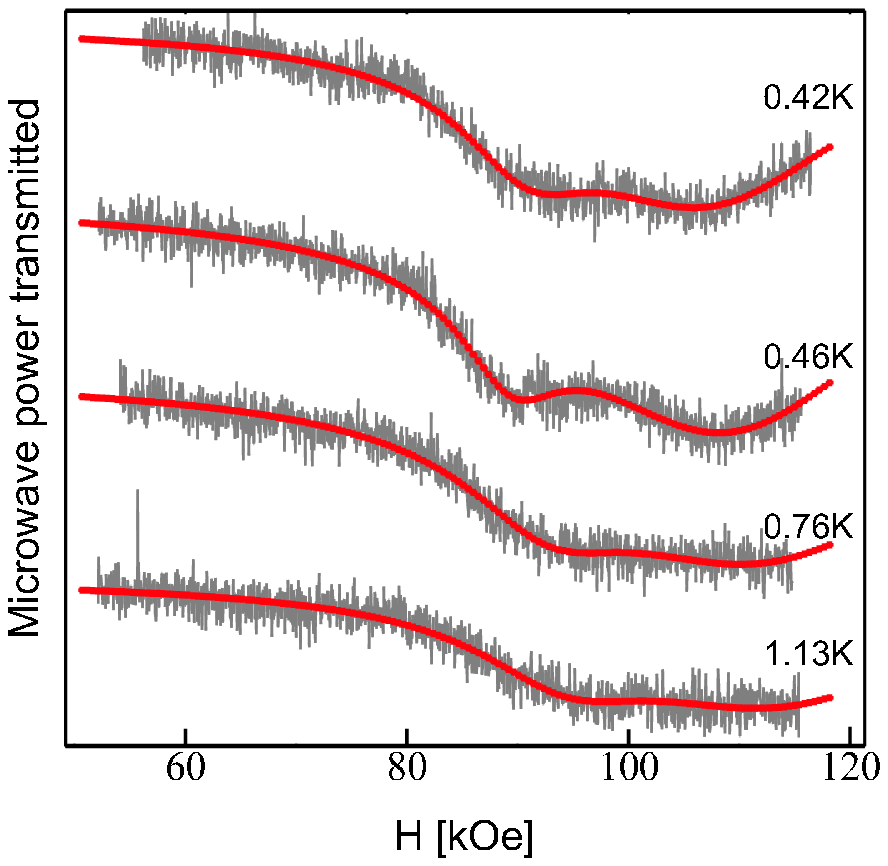, width=\figwidth, clip=}\\
  \caption{(color online) Temperature evolution of low-temperature
  modes through the N\'{e}el temperature at $f=$28.25 GHz,
  $\vect{H}||a$. Gray curves --- experiment, thick curves (red) ---
  guides to the eye obtained as a two-lorentzian best fit.} \label{fig:transitionthroughTN}
\end{figure}

Upon heating above 1K, the resonances LT1 and LT2 are no longer
resolved.  At several frequencies around 80---100 GHz we observed an
indication of weak ESR mode  at a temperature of 1.2K. This mode is,
probably, related to modes LT1 and LT2. It is not split and located
between LT1 and LT2 modes. A weak trace of the critical region
survives to about 2.5~K in the form of a wide resonance, see Figure
\ref{fig:35ghzvartatlowT}. We do not observe a clear anomaly in the
temperature evolution of LT3 and LT4 modes at the N\'{e}el
temperature (which is 0.6~K at the field of 100 kOe
\cite{hagiwara-prl96,TateiwaPhysicaB}). Instead, these modes merge
with the wide absorption near the critical field. This
transformation is shown on Figure \ref{fig:transitionthroughTN}.

\subsection{Frequency-field diagrams.}


\begin{figure}
\epsfig{file=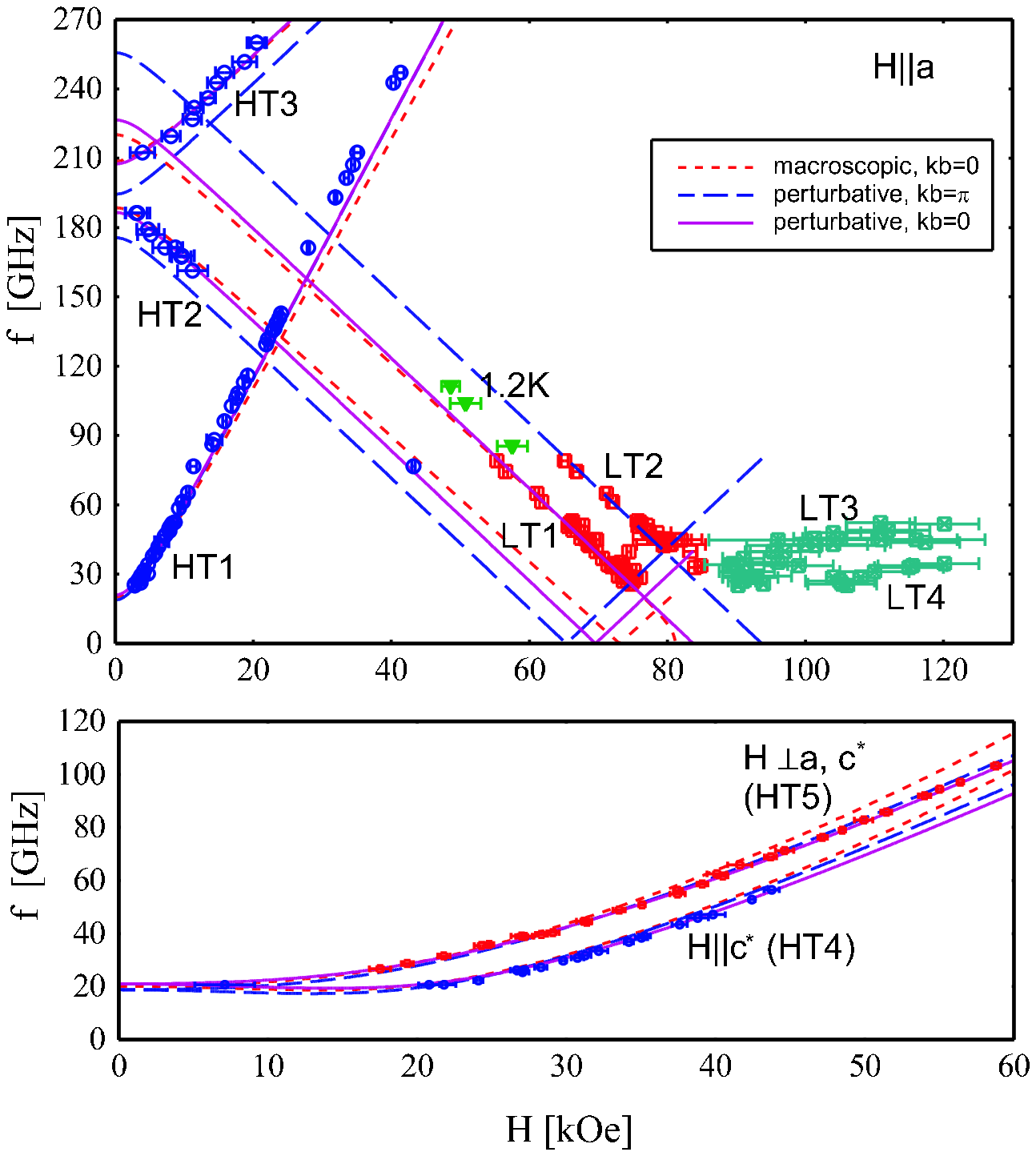, width=\figwidth, clip=}\\
\caption{(color online) Frequency-field diagram, showing all
observed resonance modes (symbols).  Curves show different model
approaches discussed in the text. Dotted line : macroscopic
(bosonic) model. Solid and dashed lines: perturbative (fermionic)
model in different points of $\vect{k}$-space. }\label{fig:compare}
\end{figure}


\begin{figure}
  \epsfig{file=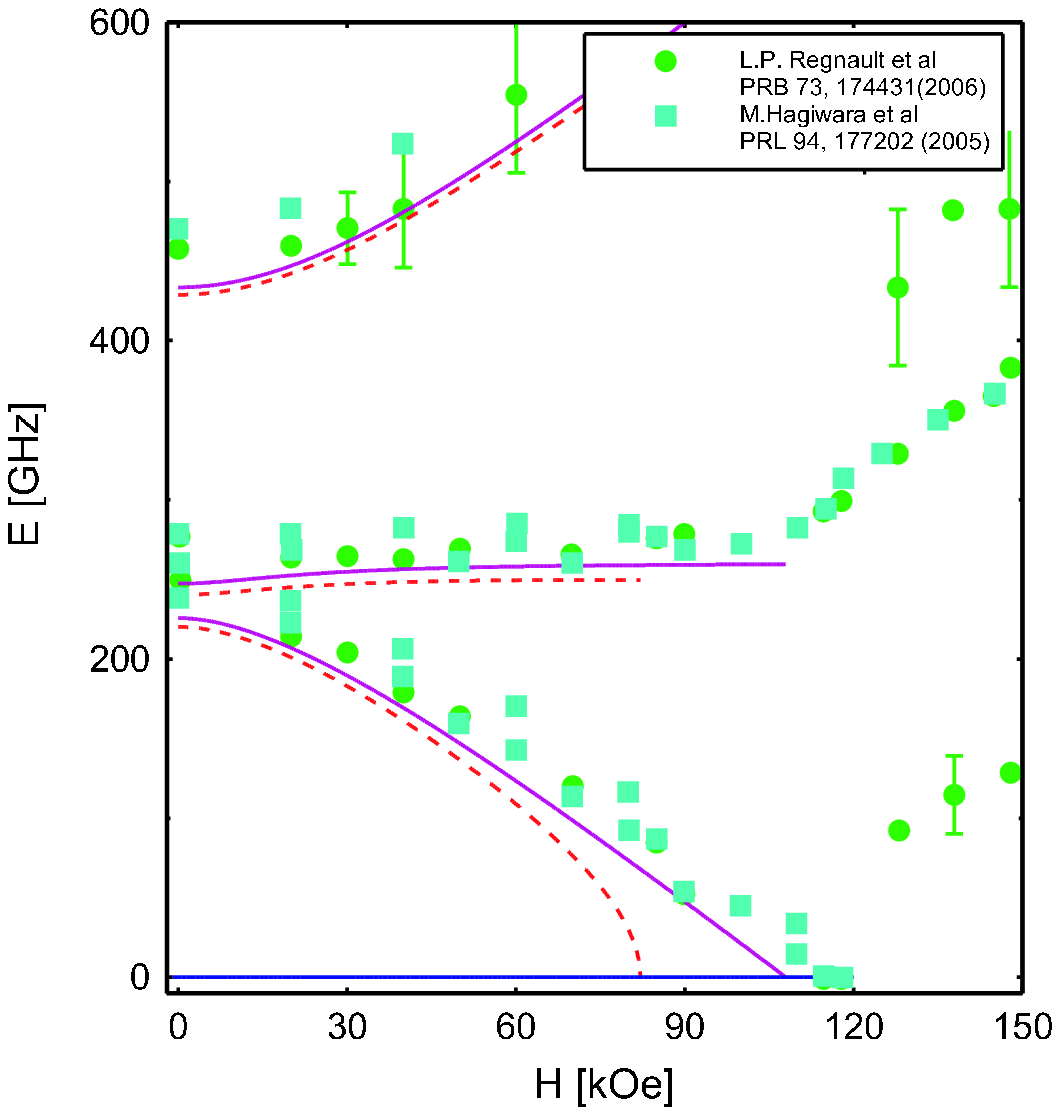, width=\figwidth, clip=}\\
  \caption{(color online) Neutron scattering data on
  the field dependence of triplet sublevels' energies from
  Refs.\onlinecite{hagiwara-prl94}, \onlinecite{zheludev2006}. Curves: predictions of the perturbative model
  (solid) and macroscopic (dashed) model. $\vect{H}||b$}\label{fig:compareneutrons}
\end{figure}

Resonance fields measured in different orientations are collected on
the frequency-field diagram  (Figure \ref{fig:compare}). We observe
three  thermally activated modes in $\vect{H}||a$ orientation (HT1,
HT2, HT3), one thermally activated mode at $\vect{H}||c^*$  (HT4)
and one thermally activated mode at $\vect{H}\perp a,c^*$ (HT5).
Positions of these modes are shown as measured at 4.2 K. At $T=$0.4
K, in the $\vect{H}||a$ orientation, we observe four modes (LT1,
LT2, LT3, LT4). The frequencies of these modes are shown as measured
at 0.4 K. A weak mode measured at 1.2 K originates, probably,  from
the temperature evolution of modes LT1, LT2 and is also plotted
here. The splitting of modes HT4 and HT5 described above is too
small to be resolvable on the Figure \ref{fig:compare}, being within
the error bars.

Thermally activated modes HT1, HT2 and HT3 demonstrate zero-field
gaps of $20\pm3$, $190\pm2$ and $210\pm2$ GHz correspondingly. The
frequencies of LT1 and LT2 modes may be approximated with a linear
law $f=-\gamma H+\Delta$ with $\gamma=2.8$GHz/kOe (assuming
$g=2.00$) and a gap $\Delta$ of $235\pm5$ and $265\pm8$ GHz for
modes LT1 and LT2 respectively. Critical fields for these modes,
estimated as $H_c=\Delta/\gamma$, are $84\pm2$ and $94\pm3$ kOe
respectively.

\section{Discussion.}
This section is organized as follows: First, we will briefly
summarize our experimental data and compare them with earlier
results. Second, we will check the correspondence of perturbative
and macroscopic approaches of spectrum description  to experimental
data. Finally, we will estimate the inter-chain coupling from the
observed splitting of resonance modes and will analyze possible
influence of magnon  decay on the spectrum terminating in NTENP.

\subsection{Qualitative analysis.}

Three types of spin resonance modes were observed: thermally
activated intra-triplet transitions, singlet-triplet transitions and
antiferromagnetic resonance absorption.

The zero field singlet-triplet transition frequencies estimated from
the linear approximation of modes LT1 and LT2 ($235\pm5$ and
$265\pm8$ GHz, correspondingly) are in a good agreement with the
lowest triplet excitation observed in neutron scattering
experiments\cite{hagiwara-prl94} at $235\pm 5$GHz (i.e.
$0.97\pm0.02$meV).  The frequencies of two other excitations were
determined from neutron scattering as $273\pm5$ and $484\pm24$ GHz
($1.13\pm0.02$ and $2.0\pm0.1$ meV). The zero-field ESR intratriplet
transitions  at $20\pm3$, $190\pm2$ and $210\pm2$ GHz correspond
well to the triplet sublevels energy differences, measured in
neutron scattering experiments as $17\pm7$, $211\pm25$ and
$249\pm25$ GHz.

The derivative $\frac{df}{dH}$ for the HT1 mode is approximately
twice as large as expected for a free spins resonance with a
$g$-factor of 2.00. Besides, the zero field frequency of the HT1
mode is much smaller than the splitting of the triplet levels due to
the main anisotropy term. This feature is well known for spin $S=1$
as a "two-quantum" transition between the states with $S_z=+1$ and
$S_z=-1$ (see, e.g. Ref.{\onlinecite{Abragam}). This type of the
$f(H)$-dependence should be observed for the direction of magnetic
field along the main anisotropy axis and is very sensitive to the
field orientation. Thus, our observation of HT1 mode proves that $a$
is close to the main anisotropy axis.

The extrapolation of modes LT1 and LT2 to zero frequency results in
the critical fields of $84\pm2$ and $94\pm3$ kOe. This is again in
agreement with the magnetization measurements\cite{narumi-prb} on
non-deuterated sample of NTENP done at 1.3 K where  the increase of
magnetization starts at 84$\pm$1 kOe for ${\bf H}\parallel a$. A
higher value of the critical field $H_c(\vect{H}||a)=91.7\pm0.1$kOe
was reported \cite{hagiwara-prl96} for deuterated NTENP. This value
was extracted from the critical exponent fit of the phase boundary
of the field-induced ordering. However, due to the lack of the
experimental data points on the phase diagram in the immediate
vicinity of the critical field this estimate of the $H_c$ should be
considered with a higher error bar.

The singlet-triplet transitions (mode LT1 and LT2) should be
forbidden in the Heisenberg case as transitions between different
spin multiplets.  "Two-quantum" transitions (mode HT1) are  also
forbidden in the case of axial anisotropy, because there is no
dipolar matrix element connecting states with $S_z=1$ and $S_z=-1$.
In the presence of in-plane anisotropy it becomes allowed and should
be observable at microwave magnetic field polarized parallel to
static field. The modes HT2 and HT3 correspond to allowed
transitions with large matrix elements at a conventional ESR
polarization ( $\Delta S_z=\pm 1$). Despite this, the modes  HT2,3
have weaker intensities and are broader than "forbidden" modes. The
intensity and lineshape for different modes will be analyzed in
Subsection D.

Finally, there is a splitting of certain resonance modes: (i)
Thermally activated modes HT4 and HT5 demonstrate splitting into two
components of approximately equal intensity. (ii) Instead of a
single-mode singlet-triplet transition, we observe  two modes LT1
and LT2, LT1 is usually much weaker. This strong difference of LT1,2
modes intensities was repeated regularly for several samples. This
excludes possible explanation of the splitting being caused by
crystallites of different orientations.

\subsection{Comparison of the results of perturbative (fermionic) and
macroscopic (bosonic) approaches to the low-energy dynamics of NTENP}

\begin{table}[h]
\caption{Comparison of perturbative (fermionic) and macroscopic
(bosonic) models with the known experimental results. Parameters of
the models are obtained from the fits to the ESR data presented in
Figure \ref{fig:compare}. \label{tab:compare}}
\begin{tabular}{|r|c|c|c|}
    \hline
        &\multicolumn{2}{|c|}{Our experiment:}&Other\\
        \cline{2-3}
        &perturbative&macroscopic&experiments,\\
        &model&model&Ref.\\
    \hline
    zero-field&&&\\
    energies, GHz&227, 247, 434 &220, 240, 429&235, 273, 484\cite{hagiwara-prl94}\\
    &&&\\
    $H_c$, kOe:&&&\\
    $\vect{H||a}$&83&81&$\approx$85\cite{narumi-prb}; 92\cite{hagiwara-prl96}\\
    $\vect{H||c^*}$&119&86&\\
    $\vect{H\perp a,c^*}$&117&83&\\
    $\vect{H||b}$&108&82&114\cite{hagiwara-prl94}\\
    \hline
\end{tabular}
\end{table}

We fit the observed ESR frequencies to the perturbative and
macroscopic models mentioned in the Introduction. The
$\vect{k}$-dependence of model parameters  will be neglected in this
Subsection because at low temperatures the  quasiparticles are
excited at the bottom of the spectrum. Consideration of the
$\vect{k}$-dependence of energy levels near the bottom of the
spectrum results in the asymmetric distortion of the absorption
line, while the maximum of absorption still corresponds to the
bottom of the spectrum due to a large statistical weight of the
corresponding states\cite{affleck}.  The main anisotropy axis $z$ is
taken to be parallel to the chain direction $a$, this choice is
based on the observation of the "two-quantum" transition mode HT1.
The second anisotropy axis may not be chosen from the symmetry
consideration of the triclinic crystal and should be derived from
the experiment. Thus, the direction of the second anisotropy axes in
a plane perpendicular to $a$ is fitting parameter of the model.
Finally, we will assume $g$-factor to be isotropic.

The perturbative model\cite{zaliznyak,golinelly} treats triplet
excitations as non-interacting S=1 quasiparticles in the effective
crystal field and in the external magnetic field. This approach is
limited to low fields when the Zeeman energy is small in comparison
with the spin gap. However, as it was shown in
Ref.\onlinecite{zaliznyak}, in a one dimensional case the results of
perturbative formalism coincides with the fermionic formalism of
Ref.\onlinecite{tsvelik}, which is valid at all fields. For this
reason, taking into account the strongly one-dimensional character
of spin interactions in NTENP, we will apply the results of the
perturbative approach all the way up to the critical field.

The effective Hamiltonian of the perturbative model is:

\begin{eqnarray}
  \ham_{tripl}&=&\Delta+D\left(\sz\right)^2+E\left(\left(\sx\right)^2-\left(\sy\right)^2\right)+\nonumber\\
  &&+g\mu_B(\vect{H}\svect)
  \label{eqn:ham}
\end{eqnarray}

\noindent here $\Delta$ is the gap separating singlet state with
zero energy from the excited triplets, $D$ and $E$ are the effective
anisotropy constants, and $g$ is the $g$-factor. This model includes
5 parameters: $\Delta$, $D$, $E$, $g$  and an angle $\phi$ between
anisotropy $x$ axis and $c^*$ direction. Given these parameters,
eigenenergies of the Hamiltonian (\ref{eqn:ham}) can be found in any
orientation of magnetic field. Differences between the eigenenergies
correspond to frequencies of thermally activated modes. The lowest
eigenenergy corresponds to the singlet-triplet transition. The
zero-field energies for this model are $\Delta$, $\Delta+D+E$ and
$\Delta+D-E$. The field dependences can be easily found for
$\vect{H}||z$:

\begin{eqnarray}
E_{1,2}&=&\Delta+D\pm\sqrt{\left(g\mu_BH\right)^2+E^2}
\label{eqn:pert-e1}\\
 E_3&=&\Delta \label{eqn:pert-e2}
\end{eqnarray}

For an arbitrary direction an analytical solution becomes too
cumbersome. Critical fields for $\vect{H}||z$ and $\vect{H}\perp z$
are

\begin{eqnarray}
  \left(g\mu_BH_{c||}\right)^2&=&\left(\Delta+D\right)^2-E^2\\
  \left(g\mu_BH_{c\perp}\right)^2&=&\Delta\frac{\left(\Delta+D\right)^2-E^2}{\Delta+D-E\cos2\alpha}
\end{eqnarray}

\noindent here $\alpha$ is the angle in $(xy)$ plane counted from
$x$ direction.

The model was fit to the experimental data for thermally activated
intertriplet transitions (modes HT1,~2,~3,~4,~5) and for LT1 as a
singlet-triplet transition.The best fit corresponds to the following
model parameters: $D=-197\pm5$ GHz, $E=10.5\pm1.0$ GHz,
$\Delta=434\pm 5$ GHz, $g=2.02\pm0.05$ and
$\phi=(0.59\pm0.04)$rad$=(34\pm2)^\circ$. The zero-field energies of
triplet sublevels in frequency units, as predicted by this model,
are collected in  Table \ref{tab:compare}. The modelled
frequency-field dependences are presented in  Figure
\ref{fig:compare} by solid lines. They give a good approximation for
the observed resonance modes.

The  macroscopic (or bosonic) approach\cite{affleck,farmar}
describes the low energy spin dynamic of the spin-gap magnet as
oscillations of a vector field $\boldeta$. This model assumes that
the spin oscillation frequencies are small with respect to the
exchange frequency and that antiferromagnetic ordering appears above
the critical field. Thus, the model requires non-zero inter-chain
coupling.

The Lagrangian of the macroscopic model (without gradient terms) is

\begin{eqnarray} \label{eqn:lagr}
L&=&\frac{1}{2}(\dot{\boldeta}+\gamma[\boldeta\times\vect{H}])^2
-\frac{A}{2}\boldeta^2+\nonumber\\
&&+\frac{b_1}{2}(\eta_x^2 + \eta_y^2 -
2\eta_z^2)+\frac{b_2}{2}(\eta_x^2-\eta_y^2).
\end{eqnarray}

\noindent here $A$ is the exchange constant describing the energy
gap, $b_1$ and $b_2$ are the effective anisotropy constants,
$\gamma$ is a gyromagnetic ratio ($\gamma=(g/2.0)\gamma_0$, where
$\gamma_0=2.80$GHz/kOe is a free electron gyromagnetic ratio). Below
$H_c$, the corresponding dynamic equation is:

\begin{eqnarray}
\ddot{\boldeta} +2
\gamma[\dot{\boldeta}\times\vect{H}]-\gamma^2H^2\boldeta+
\gamma^2\vect{H}(\boldeta\cdot\vect{H})+A\boldeta- \nonumber
\\ -b_1 \left(
    \begin{array}{c}
    \eta_x\\ \eta_y\\ -2\eta_z\\
    \end{array}
    \right)
-b_2 \left(
    \begin{array}{c}
    \eta_x\\ -\eta_y\\ 0\\
    \end{array}
    \right)  =0.\label{eqn:dyneqn}
\end{eqnarray}

For the case of $\vect{H}||z$ the excitation energies are:

\begin{eqnarray}
    E_{1,2}^2&=&\gamma^2H^2+A-b_1\pm\nonumber\\
    &&\pm\sqrt{4\gamma^2H^2(A-b_1)+b_2^2}\\
    E_3^2&=&A+2b_1
\end{eqnarray}

The critical fields for $\vect{H}||z$ and $\vect{H}\perp z$ are

\begin{eqnarray}
    \gamma^2H_{c||}^2&=&A-b_1+|b_2|\\
    \gamma^2H_{c\perp}^2&=&\min\left\{A+2b_1;\frac{\left(A-b_1\right)^2-b_2^2}{A-b_1-b_2\cos2\alpha}\right\}
\end{eqnarray}

Note, that for $b_1>0$ (i.e. $E_{1,2}<E_3$, which corresponds to the
case of NTENP) this model predicts no anisotropy of critical field
for $b_2=0$.

This model also has 5 fit parameters: $A$, $b_1$, $b_2$, $\gamma$
and the orientation of the $x$ axis with respect to the $c^*$
direction $\phi$. The best fit obtained is shown on the Figure
\ref{fig:compare}. It corresponds to the following parameters:
$A=96700\pm2000$GHz$^2$, $b_1=43600\pm1000$GHz$^2$,
$b_2=4640\pm500$GHz$^2$, $\gamma=2.71\pm0.10$GHz/kOe (i.e.
$g=1.94$), $\phi=(0.61\pm0.08)$rad$=(35\pm4)^\circ$. The critical
fields and zero-field energies of triplet states,  predicted by this
model, are collected in Table \ref{tab:compare}. This macroscopic
model also yields predictions close to the experimental data, but
the agreement is  less satisfactory. In particular, systematic
deviations are found for modes HT1,~4,~5.

We conclude that the ESR data (Figure \ref{fig:compare}) favours the
perturbative model though not unambiguously. To provide an
additional check, we have calculated field dependences of energy
levels in the $\vect{H}||b$ orientation of magnetic field using the
model parameters values determined above. These model predictions
can be compared with independent data from neutron scattering,
obtained in the same orientation of magnetic field
\cite{hagiwara-prl94, zheludev2006}. As born out in Figure
\ref{fig:compareneutrons}, predictions of the macroscopic model
visibly deviate from the experimental points, while the perturbative
model describe the neutron scattering data fairly well.

Thus we conclude that the perturbative (fermionic) model gives a
better correspondence with the experiments for the case of NTENP.
Disagreement of the macroscopic model with the experiment is
probably due to the violation of  low-frequency condition, since the
energy of the upper triplet component of 450 GHz is about 40\% of
the largest exchange $J_2/h$.

\subsection{The effect of inter-chain interaction.}

Weak inter-chain interactions estimated as $J_\perp\sim
0.01$meV$\approx0.1$K were reported in
Ref.\onlinecite{zheludev2004}. These interactions will result in a
dispersion of excitations propagating transverse to the chains. To
take into account inter-chain coupling in the model calculations we
will use form of excitations spectrum that combines the energy gaps
and dispersive terms  analogously to an antiferromagnet with
anisotropic spin wave velocities:

\begin{eqnarray}
E_i^2&=&\Delta_i^2+s_{\parallel}^2\sin^2(\vect{k}\vect{a})+s_{\perp}^2\sin^2(\vect{k}\vect{b})+\nonumber\\
&&+2s_{\parallel}s_{\perp}(1-\cos(\vect{k}\vect{a})\cos(\vect{k}\vect{b}))\label{eqn:k-spectrum}
\end{eqnarray}

\noindent here $s_{||,\perp}=2SJ_{||,\perp}$ are spin-wave
stiffnesses along the chain and perpendicular to the chain,
correspondingly (we use exchange energy in the form $E=\sum_{\langle
i,j\rangle} J_{ij}\vect{S}_i\vect{S}_j$), $\vect{a}$ and $\vect{b}$
are translation vectors of the rectangular lattice along the chains
and in the transverse direction respectively, $\vect{k}$ is the
vector in $\vect{k}$-space relative to the point of minimum of
energy, and $\Delta_i$ ($i=1,2,3$) are the corresponding gaps,
respectively.

The longitudinal spin-waves stiffness $s_{||}$ is due to the
dominating in-chain exchange interaction and has a value of
\cite{zheludev2004}: $s_{||}=8.6$meV$=2080$GHz.  For a weak
inter-chain interaction $s_{\perp}\ll s_{||}$,  and the dispersion
for excitations propagating along the $b$-axis is:

\begin{eqnarray}
E_i|_{\vect{k}\vect{a}=0}&\approx&\sqrt{\Delta_i^2+2s_{\parallel}s_{\perp}(1-\cos(\vect{k}\vect{b}))}\approx\nonumber\\
&\approx&\Delta_i+\frac{s_{||}s_{\perp}}{\Delta_i}(1-\cos(\vect{k}\vect{b}))\label{eqn:profile}
\end{eqnarray}

From the exchange constants obtained in neutron scattering
experiments we have $\frac{s_{\parallel}s_{\perp}}{k_B\Delta_i}\sim
1.6$K. Thus, all excitations with transverse wavevectors should be
populated above 1K, including those at $\vect{k}\vect{b}=\pi$.
Therefore, both points with maximum density of states of excitations
will contribute to microwave absorption: the bottom of the spectrum
at $\vect{k}\vect{b}=0$ and the saddle point at
$\vect{k}\vect{b}=\pi$. The parameters of effective Hamiltonian
(\ref{eqn:ham}) in these points ($\Delta_0$, $D_0$, $E_0$ and
$\Delta_{\pi}$, $D_{\pi}$, $E_{\pi}$, respectively) are slightly
different. This will result in the splitting of the resonance modes.

The two sets of effective Hamiltonian parameters are related as
follows (see Eqns. (\ref{eqn:pert-e1}), (\ref{eqn:pert-e2}),
(\ref{eqn:profile})):

\begin{eqnarray}
    \Delta_{\pi}&=&\sqrt{\Delta_0^2+4s_{||}s_{\perp}}\\
    E_{\pi}&=&\frac{\sqrt{(\Delta_0+D_0+E_0)^2+4s_{||}s_{\perp}}}{2}-\nonumber\\
    &&-\frac{\sqrt{(\Delta_0+D_0-E_0)^2+4s_{||}s_{\perp}}}{2}\\
    D_{\pi}&=&\frac{\sqrt{(\Delta_0+D_0+E_0)^2+4s_{||}s_{\perp}}}{2}+\nonumber\\
    &&+\frac{\sqrt{(\Delta_0+D_0-E_0)^2+4s_{||}s_{\perp}}}{2}-\Delta_{\pi}
\end{eqnarray}

Now we can ascribe LT1 and LT2 modes to the transitions between the
ground state and $\vect{kb}=0$ and $\vect{kb}=\pi$ points in the
excitations spectrum. This allows us to extract the value of
$(s_{||}s_{\perp})$ from the splitting of these modes. The best fit
corresponds to $\sqrt{s_{||}s_{\perp}}=60\pm5$GHz (0.25meV).

The inter-chain exchange integral can be estimated as follows:
$J_{\perp}=(s_{\perp}/s_{||})J_{||}=((s_\perp
s_{||})/s_{||}^2)J_{||}\sim0.05$~K. This value is quite small and
there are other interactions (i.e. dipolar, or anisotropic exchange
interactions) of the same strength. Thus, the appearance of a weak
transverse dispersion may be due to the combined action of several
interactions, which are much weaker than the dominating in-chain
exchange.

Frequency-field dependences for both the $\vect{k}\vect{b}=0$ and
$\vect{k}\vect{b}=\pi$ excitations, calculated for all modes, are
shown for comparison on the Figure \ref{fig:compare}.  The splitting
of resonance lines are well described by our  dispersive model: The
splitting is largest for the singlet-triplet transition (modes LT1
and LT2). It is due to the different values of lowest triplet
sublevel energy in different points of $\mathbf{k}$-space. For the
thermally activated modes, the splitting is much weaker, since it is
determined by the difference of the triplet sublevels energies and
all sublevels are shifted in the same direction due to the
dispersion. For the HT1 mode the splitting is not observable because
of the steep increase of frequency with magnetic field, while for
modes HT4 and HT5 a weaker frequency-field dependence allows us to
resolve the doublet by scanning the field. The predicted splitting
of modes HT4 and HT5  is about 1kOe which is in agreement with
experimental observations. As for the high-frequency modes HT2 and
HT3 ($\vect{H}||a$), the predicted splitting of approximately 5kOe
may not be observed because of the large linewidths of these modes
(about 5kOe).

Equal  intensities of the components of the split doublet for
thermally activated modes HT4 and HT5 are due to almost equal
population numbers at the top and bottom of the spectrum of
transversal excitations. The observation of both LT1 and LT2 modes
indicates that the ground state is a mixture of $\vect{kb}=0$ and
$\vect{kb}=\pi$ states.

\subsection{Indications of the quasiparticles decay.}

\begin{figure}
  \epsfig{file=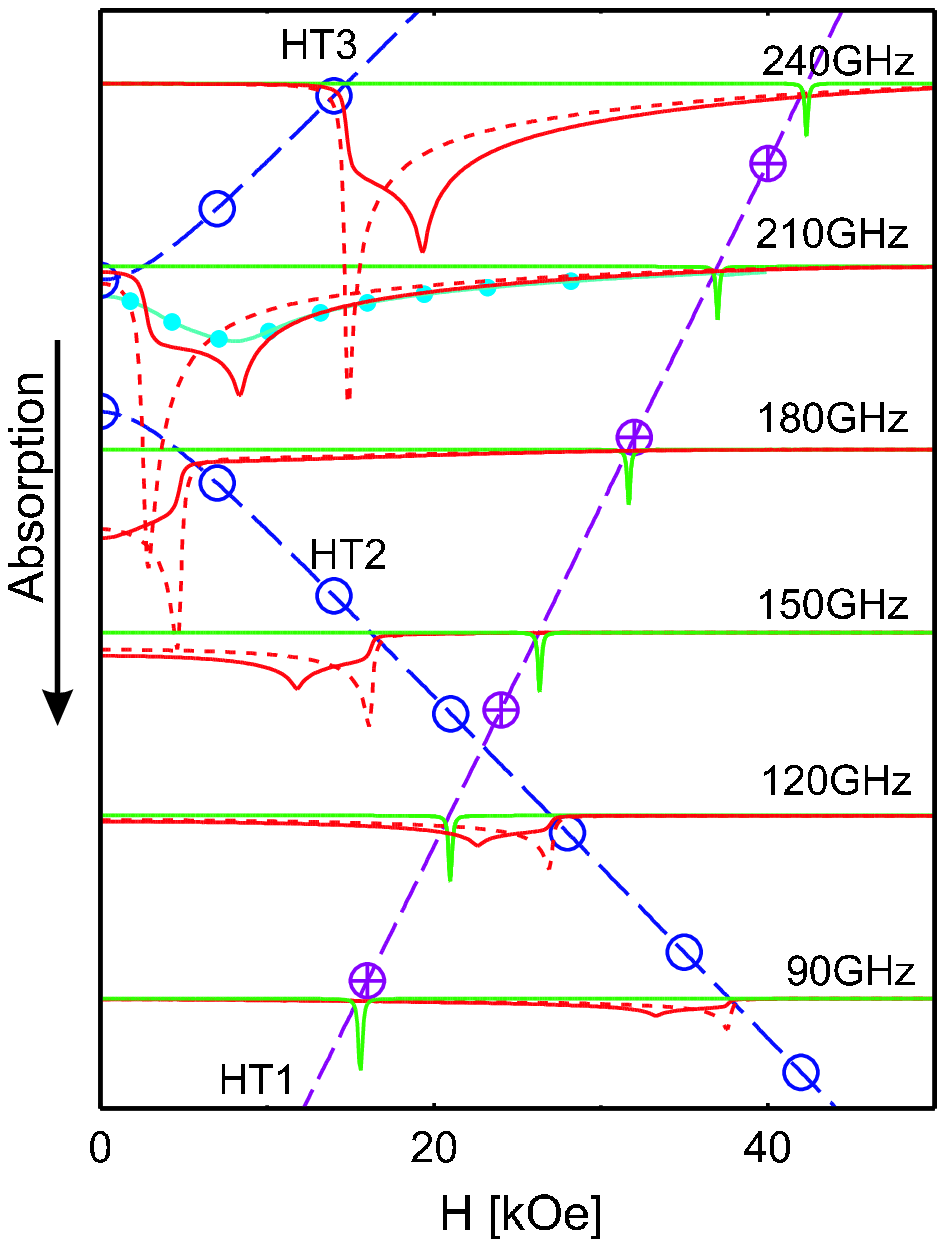,  width=\figwidth, clip=}\\
  \caption{(colour online) Modelled lineshapes of   ESR absorption at 4K for
  $\vect{H}||a$.  Solid lines: modelling taking into  account dispersion both along the chain
  and in the  transverse direction.   Dashed lines: modelling taking   into account only dispersion along
  the chains.  Solid line with symbols at 210GHz: modelling taking
  into account higher inherent linewidth of the $S_z=0$ magnon.
  Dashed lines with symbols: guides to   the eye marking position
  of different modes at different frequencies.}
  \label{fig:modelled}
\end{figure}

\begin{figure}

  \epsfig{file=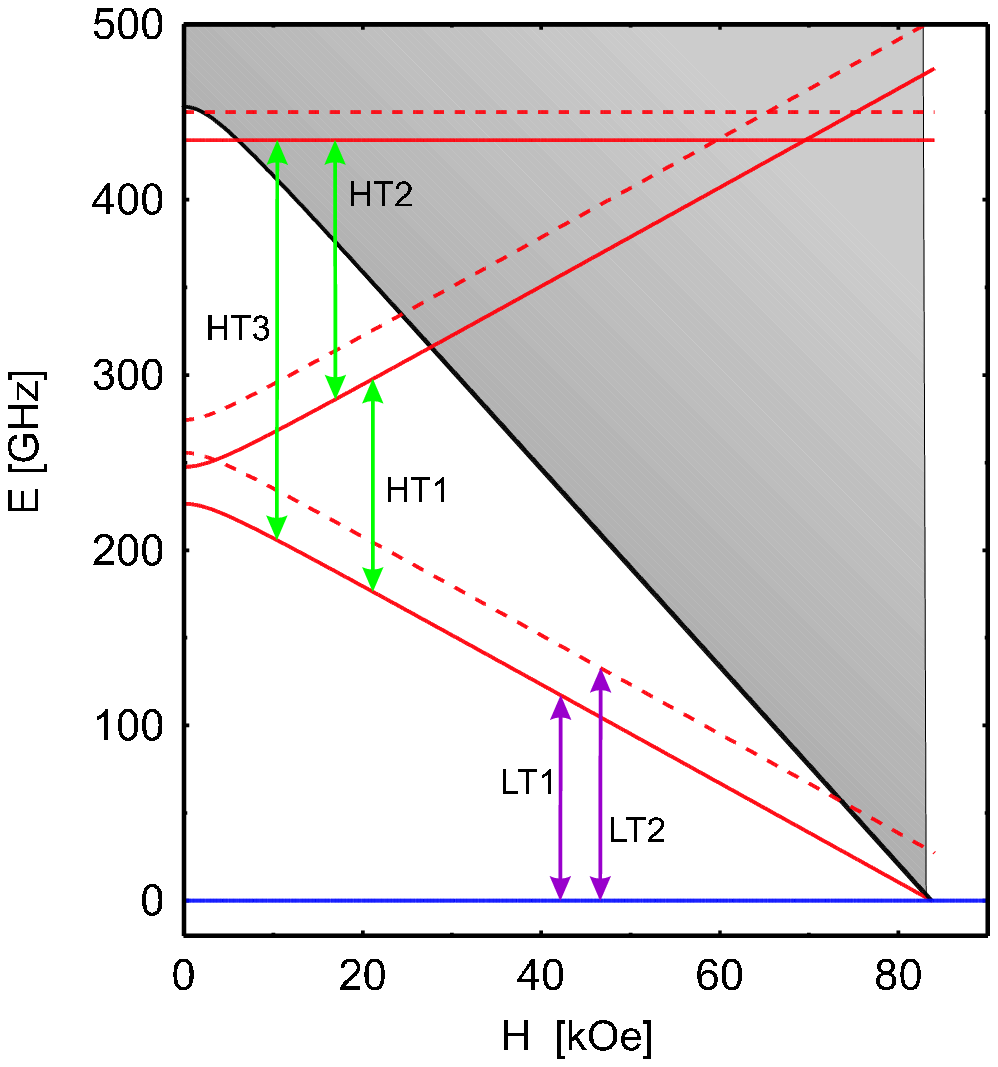, width=\figwidth, clip=}
  \caption{(colour online) Modelled field dependences of triplet energy levels
  (perturbative model). Boundaries of the zone due to the dispersion
  of excitations in the transverse directions are marked by solid
  ($\vect{k}\vect{b}=0$) and dashed ($\vect{k}\vect{b}=\pi$) lines. Arrows marks
  transitions corresponding to the observed ESR modes. For modes
  HT1...HT3 only transitions between $q_\perp b=0$ states are shown
  for the sake of simplicity. The grey shaded area shows the two-particle continuum,
  corresponding to the lowest excitation.}\label{fig:e(h)}
\end{figure}

Taking into account  the real values of the derivative $df/dH$, the
linewidth of modes HT2, HT3 is about 20 GHz, while the linewidth of
HT1 is 1 GHz. For modes HT4,5 it is about 4 GHz (splitting
included).

We performed a modelling of the ESR lineshape by integrating the
expected signals over the $\vect{k}$-space.

Model calculations were performed in the perturbative approach using
the best fit parameter values obtained as described in previous
sections. The parameters $\Delta_{\vect{k}}$, $D_{\vect{k}}$ and
$E_{\vect{k}}$ for the effective Hamiltonian (\ref{eqn:ham}) where
calculated using the zero field energies given in Table
\ref{tab:compare} for the perturbative model and using
Eqn.(\ref{eqn:k-spectrum}). Then eigenfrequencies and eigenfunctions
for Hamiltonian (\ref{eqn:ham}) and
 transition matrix elements were found. Finally,
we calculated absorption by summing over the $\vect{k}$-space,
taking into account the population of the sublevels involved in a
given transition. The lineshape of individual transition was assumed
to be Lorentzian with the inherent width of 0.5~GHz. This value was
found to reproduce well the low-temperature linewidth of the narrow
HT1 mode.   The inherent linewidth was assumed to be the same for
all transitions. The absorption was calculated for optimal
polarization of the microwave magnetic field, i.e. for HT1 mode the
longitudinal polarization of microwave field was used, while HT2 and
HT3 modes absorption was calculated at transverse polarization. In
an experiment both polarizations are usually present. For low
frequency modes, the transverse polarization dominates.

Results of this modelling are presented on Figure
\ref{fig:modelled}. The model predicts that the absorption by modes
HT2, HT3 should be in a wide band with a sharp edge corresponding to
$\vect{k}\vect{b}=0$ and a peak corresponding to
$\vect{k}\vect{b}=\pi$, instead of the wide and approximately
Lorentzian lines observed. The model predicts also, that the
resonance absorption at the HT2 mode should be large enough to be
detected at least at frequencies above 100 GHz.  It disagrees with
experimental observations, because,  in the frequency range 170-240
GHz, a smeared peak is observed experimentally. Besides, below this
frequency range, the resonance HT2 was not observed, while HT1 is
observable.

A simple way to reconcile the model with the experiment is to
suppose that  the inherent linewidth of HT2 and HT3 modes is much
larger then that for HT1 mode. The broad modes HT2 and HT3 involve
transitions to the highest excitations branch, corresponding to
$S_z=0$, while the narrow modes HT1, HT4, HT5, LT1, LT2 correspond
to transitions between lower energy branches. Therefore, the
existence of a strong relaxation specifically for the $S_z=0$
excitations could explain observed discrepancy. We model this strong
relaxation by assuming inherent linewidth of HT2 and HT3 modes to be
equal to  5~GHz instead of 0.5~GHz. This, indeed, results in the
smeared lineshape of HT3 resonance, as shown in
Fig.\ref{fig:modelled} in a solid line connecting filled circles.

Possible reason for strong relaxation for the $S_z=0$ excitations is
the entry of this branch into the two particle continuum range.
Similar explanation was proposed in Ref.\onlinecite{hagiwara-prl94}
to describe the disappearance of the highest-energy excitations
observed in neutron scattering experiment. The frequency
corresponding to the bottom of the two particle continuum may be
derived as a doubled frequency of the $\vect{k}=0$ excitation (for
the excitations with a positive dispersion). Energies of excitations
and two-magnon continuum related to the lowest branch are shown in
Figure \ref{fig:e(h)}. The decays of highest excitation branch into
two excitations from the lowest branch may occur in NTENP due to the
close values of the energy of highest excitation mode (which are 435
GHz for $\vect{k}\vect{b}=0$ and 450 GHz for $\vect{k}\vect{b}=\pi$)
and the bottom of two particle continuum (which is 453GHz for
$H=0$). In a magnetic field, the energy of the lowest branch
decreases. At fields above 6 kOe the excitations of the highest
branch became unstable with respect to the decays into two magnons
of the lower branch.

It is an open question whether this intersection with the continuum
destroys the excitation branch or just makes the magnon life time
shorter. In any case, decay should result in a broadening of ESR
line. The observed disappearance of the HT2 resonance in fields
above 20 kOe (or in the frequency range below 150 GHz) may be due to
such decays into two particles of lower branch.

\subsection{Antiferromagnetic resonance}

The antiferromagnetic resonance absorption above $H_c$ consists of a
broad band with two absorption maxima. This observation corresponds
to the reported extreme broadening of the lowest energy
antiferromagnetic mode in neutron scattering above $H_c$ for
$\vect{H}||b$. \cite{zheludev2006} The splitting of the
antiferromagnetic mode into two close resonances LT3 and LT4  may be
due to the a dispersion in the transverse direction, analogous to
the splitting of LT1,2 and HT4,5 modes described above.

\section{Conclusions.}

The low-energy spin dynamics of the dimerised S=1 spin-chain
compound NTENP is studied by means of ESR. Modes of magnetic
resonance corresponding to transition between the spin sublevels of
the collective triplet excitations of  the spin-gap magnet are
observed and identified. The singlet-triplet transitions to a spin
gap state are also observed. The data analysis demonstrates that
low-energy spin dynamics of this spin-gap magnet may be described by
a perturbative model in the whole field range up to  the critical
field. The singlet-triplet transition and the intratriplet
transitions have a doublet structure due to the dispersion of
excitations propagating perpendicular to the chains. The energy of
the inter-chain interaction is determined from the doublet
splitting.

In addition we have observed a strong broadening and termination of
spin resonance modes at the boundary of the two particle continuum,
indicating the influence of the two-magnon decay processes on the
spectrum of excitations in NTENP.

\acknowledgements

The work was supported by grants of Russian Foundation for Basic
Research (projects N 09-02-12341, 09-02-00736-a). One of the authors
(V.G.) was supported by Presidential Grant for Young Scientists
MK-4569.2008.2. Research at Oak Ridge sponsored by the Material
Sciences and Engineering Division, Office of Basic Energy Sciences,
US Department of Energy.

Authors thank M.Zhitomirsky, I.Zaliznyak and O.Petrenko for their
interest to the work and useful  discussions.

\end{document}